\documentclass[letterpaper,twocolumn,10pt]{article}
\usepackage{usenix2019_v3}

\usepackage{tikz}
\usepackage{amsmath}

\usepackage{filecontents}

\usepackage{xcolor}
\usepackage{framed}
\usepackage{multirow}
\usepackage{subfig}
\usepackage[linesnumbered,algoruled,boxed,lined]{algorithm2e}
\usepackage{pifont}
\usepackage{bbding} 

\definecolor{BrickRed}{RGB}{203,65,84} 
\definecolor{DeepGreen}{RGB}{0,100,0}
\definecolor{DeepRed}{RGB}{139,0,0}

\usepackage{amssymb}

\usepackage{enumitem}

\usepackage[misc]{ifsym} 
\usepackage{graphicx}
\usepackage{booktabs} 
\usepackage{cite}
\usepackage{threeparttable}

\usepackage{tcolorbox}
\usepackage{setspace}
\usepackage{listings}

\usepackage[normalem]{ulem}

\usepackage[switch]{lineno}
\definecolor{darkblue}{rgb}{0,0,0.55}

\newcommand{\myeat}[1]{}
\newcommand{\myparhead}[1]{\textbf{#1}.}
\newcommand{\mytechno}[1]{\texttt{#1}} 
\newcommand{\myreffig}[1]{Figure \ref{fig:#1}}
\newcommand{\myrefsec}[1]{\textsection\ref{sec:#1}}
\newcommand{\myreftab}[1]{Table \ref{tab:#1}}
\newcommand{\myrefeqn}[1]{Equation \ref{eqn:#1}}
\newcommand{\myrefalgo}[1]{Algorithm \ref{algo:#1}}
\newcommand{\myrefapp}[1]{Appendix \ref{app:#1}}
\newcommand{\myreflst}[1]{Listing \ref{lst:#1}}

\definecolor{redbg}{RGB}{254,241,240}
\definecolor{redoutline}{RGB}{252,163,152}
\definecolor{redtext}{RGB}{207,24,34}

\renewcommand\footnotemark{} 

\usepackage[available]{usenixbadges}

\begin{document}


\date{}

\title{\Large \bf I Know What You Said: Unveiling Hardware Cache Side-Channels in Local Large Language Model Inference}

\author{
Zibo Gao\textsuperscript{1,2}, Junjie Hu\textsuperscript{1,2}, Feng Guo\textsuperscript{1,2}, Yixin Zhang\textsuperscript{1,2,\textrm{\Envelope}}, Yinglong Han\textsuperscript{1,2}, Siyuan Liu\textsuperscript{1,2}, \\ Haiyang Li\textsuperscript{1,2}, and Zhiqiang Lv\textsuperscript{1,2,\textrm{\Envelope}} \\
\textsuperscript{1}Institute of Information Engineering, Chinese Academy of Sciences. \\
\textsuperscript{2}School of Cyber Security, University of Chinese Academy of Sciences. \\ 
{Email: \{gaozibo,hujunjie,guofeng,zhangyixin,hanyinglong,liusiyuan,lihaiyang,lvzhiqiang\}@iie.ac.cn}
\thanks{\Envelope~Corresponding authors.}
}

\maketitle

\begin{abstract}

Large Language Models (LLMs) that can be deployed locally have recently gained popularity for privacy-sensitive tasks, with companies such as Meta, Google, and Intel playing significant roles in their development.
However, the security of local LLMs through the lens of hardware cache side-channels remains unexplored.
In this paper, we unveil novel side-channel vulnerabilities in local LLM inference: token value and token position leakage, which can expose both the victim's input and output text, thereby compromising user privacy.
Specifically, we found that adversaries can infer the token values from the cache access patterns of the token embedding operation, and deduce the token positions from the timing of autoregressive decoding phases.
To demonstrate the potential of these leaks, we design a novel eavesdropping attack framework targeting both open-source and proprietary LLM inference systems.
The attack framework does not directly interact with the victim's LLM and can be executed without privilege.

We evaluate the attack on a range of practical local LLM deployments (e.g., Llama, Falcon, and Gemma), and the results show that our attack achieves promising accuracy.
The restored output and input text have an average edit distance of 5.2\% and 17.3\% to the ground truth, respectively.
Furthermore, the reconstructed texts achieve average cosine similarity scores of 98.7\% (input) and 98.0\% (output).
\end{abstract}
	

\section{Introduction}
Large Language Models (LLMs), such as OpenAI's ChatGPT~\cite{chatgpt}, Meta's Llama~\cite{llama}, Google's Gemma~\cite{gemma}, and Mistral AI's Mistral~\cite{mistral}, have enabled a broad spectrum of applications ranging from chatbots to personal agents.
Their ability to follow human instructions and make decisions has garnered substantial attention from the public, reshaping the digital landscape.

However, users may unintentionally disclose private data while interacting with LLMs.
For instance, Samsung Electronics exposed its confidential data to a cloud-based LLM service in 2023, resulting in a ban on employee use of generative AI tools~\cite{Samsungevent}.
To mitigate the risk of sharing sensitive data with third parties, locally deployed LLMs have garnered increasing attention for their suitability in privacy-critical tasks~\cite{GPT4All,edgefm-ondevllm,10.1145/3675094.3677545,10.1145/3669940.3707239}. This paradigm is supported by corporations such as Meta, Google, and Microsoft.
Local LLMs are suited for handling sensitive tasks such as editing confidential emails, seeking advice on personal matters, and assisting with financial analysis~\cite{GPT4All,financial-llm}.

Unfortunately, we find a new security threat in which LLMs' sensitive input and output text during inference can be leaked through hardware cache side channels, which has not been previously reported, to the best of our knowledge.
Specifically, we shed light on several fundamental characteristics of LLM inference that cause the side-channel leakage:
(\romannumeral1) LLMs depend on \emph{token embedding}~\cite{attn-is-all-you-need,gu2024-mamba,lieber2024-jamba,devlin-etal-2019-bert,yang-2019-xlnet}, which is essential for converting text into semantic representations that the model can process.
However, this embedding operation creates secret-dependent data access, exposing the input's token (akin to words) values through cache access patterns.
(\romannumeral2) LLM inference generally follows the \emph{autoregressive} paradigm, where each output token is recursively fed into the model and passed through the token embedding, enabling the side-channel leakage for both the input and output tokens.
Moreover, the autoregressive generation is inherently sequential and unfolds over multiple time steps, implying that the timing of embedding operations correlates with the position of the output token.
Exploiting these characteristics, adversaries can launch a spy application co-located with the LLM to probe cache access timing and form cache traces, then map the cache traces to the victim's input and output text.

\myparhead{Challenges} We need to overcome several unique challenges to realize the attack in practical scenarios.
First, the cache side-channel exhibits a relatively low signal-to-noise ratio (SNR) due to system activities during LLM inference.
This noise can result in missing or randomly valued words in the resulting text.
Second, in practical systems, the time order of the token embedding operation for the model input is interleaved or overlapped in the time axis of the cache trace, resulting in a shuffled order of the mapped input tokens.
The root cause lies in the batching of input tokens.
Their batched processing through parallel computing results in closely clustered execution times and an unpredictable sequence of operations.

\myparhead{Solution} To address the first challenge, we propose a novel text reconstruction algorithm that fuses both the \emph{timing signal} and \emph{token list} mapped from the cache trace.
Specifically, we utilize Power Spectral Density (PSD) in signal processing to analyze the trace's timing signal (time series of the cache hit events).
We find that the timing signal during the LLM decode phases exhibits strong periodicity, whereas false positive noise shows randomness with an evenly distributed spectrum.
Leveraging these findings, we synthesize a dataset to train LLMs to capture the timing patterns and reconstruct the clean text.
During dataset synthesis, we sample tokens from publicly available textual corpora and assign each token a periodic timestamp to simulate the timing signal.
Then, we remove a random subset of the tokens to simulate false negative noise, and insert random tokens at uniformly distributed time points to simulate false positive noise.
We then fine-tune LLMs on the synthesized dataset to reduce noise and reconstruct the victim mode's output text from the nosy cache trace.

To address the second challenge, we exploit the contextual dependence between model input and output text.
Specifically, we fuse the token list mapped from the cache trace and the reconstructed output text as a whole context, then fine-tune LLMs on a randomly synthesized dataset to restore the original response from the context.

After overcoming the above challenges, we present the first LLM-targeted hardware cache side-channel eavesdropping framework.
The attack does not directly interact with the victim model.
Instead, it only passively (i.e., being stealthy) observes the behavior of the hardware cache shared with the LLM inference process and does not rely on profiling the victim.

\myparhead{Evaluations}
We demonstrate the feasibility of the eavesdropping attack on a range of popular LLMs (e.g., Meta Llama, Google Gemma, and Microsoft Phi) deployed on a variety of popular LLM frameworks, such as HuggingFace transformers, Intel IPEX-LLM, open-source llama.cpp.
Our empirical experiments show that the attack can effectively restore the full text of the model output and reconstruct the model input with approximate semantics.
When attacking llama.cpp, our method achieved an average Levenshtein similarity (1 - Edit Distance) of 94.8\% and 82.7\% for the restored output and input text, respectively.
Furthermore, our method reached an average cosine similarity score of 98.7\% (output) and 98.0\% (input), demonstrating significant information leakage.


\myparhead{Contributions}
In a nutshell, the contributions of this paper include:
\begin{itemize}
\item \emph{Novel Side-Channel Attack.}  We introduce a new side-channel attack against LLM inference, which leaks both the model input and output text via hardware cache side-channels.

\item \emph{New End-to-end Attack Framework.}
We present the first hardware cache side-channel attack framework targeted both model input and output of various currently deployed LLMs, without profiling or directly interacting with the victim model.

\item \emph{Efficient Dataset Synthesis Strategy.} We innovate in automatically synthesizing datasets and train LLMs to capture the timing patterns of the cache trace and reconstruct the model output text from the noisy cache trace.

\item \emph{New Input Reconstruction Strategy.} We fine-tune LLMs to reconstruct the original order of shuffled input tokens via the contextual dependence between model input and output text.

\item \emph{Empirical Study.}
We conduct experiments in real-world scenarios to evaluate the threat of the proposed attack strategies.
Additionally, we discuss mitigation of the attack and security recommendations for LLM inference framework implementations.
\end{itemize}

\section{Background}
\subsection{Large Language Model Inference}
\label{sec:llm_inference}

An LLM is typically a generative model that estimates the probability distribution of the next token (akin to words) conditioned on the given model input and context.
Based on the token distribution, LLMs sample new tokens to generate model output.

\myparhead{Token and Tokenization} A token is the smallest meaningful unit of natural language, which can be a word or a part of the word.
Tokens are generated by a process named tokenization, which segments a text into a sequence of individual units denoted by $[t_1, t_2, ..., t_N]$, where for each $1\le i\le N, t_i \in V$, and $V$ is the vocabulary.
For example, a text tokenized by a Byte Pair Encoding (BPE) tokenizer is shown as follows:
\begin{center}
$['\mathrm{The}','\mathrm{ quick}','\mathrm{ fox}', '\mathrm{ jumps}', '\mathrm{ over}', '\mathrm{ the}', '\mathrm{ lazzy}', '\mathrm{ log}']$
\end{center}

Similarly, the above token sequence can be mapped back to the original text through de-tokenizers, which implement the inverse function of the tokenizers.

We note that tokenizers and de-tokenizers used by different LLMs generally follow similar principles of segmenting the text into meaningful units.
Moreover, most LLM vendors do not consider the tokenizers to be secrets.
Even proprietary vendors like OpenAI have made their tokenizers public~\cite{openai-tokenizer}.

\myparhead{Model Input} The input to the LLM is a token sequence, denoted as $I=[t_1,t_2,...,t_N]$, where $N$ is the input length.
The input commonly contains system and user prompts, which can be human instructions or other requests that the user wants the LLM to respond to.

\myparhead{Model Output} In response to the input, the LLM generates an output text, also denoted as $O=[t_{N+1},t_{N+2},...t_{N+M}]$ where $M$ is the output length.

\myparhead{Model Inference}
LLM inference refers to the process where the model accepts the given input and context and generates responses, wherein the context consists of previously generated text.
Generally, LLM inference follows the autoregressive paradigm that can be divided into two phases.

In \emph{prefill phase}, LLM accepts all the input tokens in this phase.
It estimates the distribution $P(t_{N+1}|t_1,t_2,...,t_{N})$ through neural networks.
This distribution decides the first new token $t_{N+1} \sim P(t_{N+1}|t_1,t_2,...,t_{N})$.

In \emph{decode phases}, LLMs output the subsequent tokens \emph{autoregressively} one at a time until encountering an end-of-sentence (EOS) token.
The $i$-th token in the output is obtained by decoding the distribution $t_{N+i}\sim P(t_{N+i}|t_1,t_2,...,t_{N+i-1})$.
Then, the sampled token $t_{N+i}$ is fed into the model again to compute $P(t_{N+i+1}|t_1,t_2,...,t_{N+i})$ and produce the next token $t_{N+i+1}$.
Tokens must be into the numerical representation that the neural networks can process; therefore, LLMs fundamentally require the \emph{token embedding} operation to retrieve the embedding vectors for each token.

\myparhead{Local Inference} Traditionally, LLMs were deployed in expensive high-end hardware in data centers~\cite{powerinfer}.
With the evolution of LLM quantization~\cite{icml/4bitllm}
, pruning~\cite{nips/llm-pruner}, and optimizations of operators~\cite{DBLP:conf/iclr/Dao24,pagedattention}, local LLM inference has become feasible on more accessible hardware such as consumer-grade PCs.
For example, Intel has released a white paper on CPU-based LLM inferences~\cite{intelLlama2}.
Furthermore, there is a growing trend toward AI PCs~\cite{intel-ai-pcs} that deploy LLMs in low-cost personal devices.

Unlike cloud LLMs, local LLMs avoid sending private data to third parties on the Internet, therefore reducing the attack surfaces associated with remote services.
Hence, local LLMs are preferred by privacy-critical tasks~\cite{GPT4All,edgefm-ondevllm,10.1145/3675094.3677545,10.1145/3669940.3707239}.

\subsection{Cache Side-Channel Attacks}
\label{sec:cache_sca}
Modern processors extensively use caches to mitigate the high latency and inadequate bandwidth of off-chip memories (e.g., DRAMs).
When the processor loads or stores memory locations, it first attempts to access the data from the caches.
If the desired data are present in the caches (on local or remote cores), it can be directly accessed on the chip with minimal latency.
We refer to this case as \emph{cache hit}.
Conversely, if \emph{cache miss} occurs, the processors will fetch the data from the off-chip memories, which incurs latency of an order of magnitude higher than that of a cache hit.
Moreover, cache coherence protocols allow the sharing of remote cache contents across process cores and sockets within a machine.
Once attackers share memory and machine with the victim, they can probe the memory's load latency to infer whether the victim program accesses the data unit, resulting in cross-core and cross-socket side-channel attacks such as flush+reload~\cite{flush+reload}, flush+flush~\cite{flush+flush}, and invalidate+transfer~\cite{invalidate+transfer}.
The cross-core attacks apply to both same-core (with shared local caches) and different-core (with coherent caches) co-location scenarios.

\myparhead{Secret-Dependent Data Access}
Adversaries may infer confidential data of a program if secret values influence its memory access.
For example, the Rijndael Advanced Encryption Standard (AES) proposal employs several pre-computed lookup tables, namely T-box, to optimize performance~\cite{aes-proposal}.
However, without proper protections, the table-lookup implementation exposes secret-dependent data access.
By inferring which data unit the victim program visits, adversaries can ultimately recover the secret key~\cite{cache-atk-aes-sp11,cache-atk-aes-2006}.

\section{Attack Overview}
In this section, we provide an overview of the eavesdropping attack.
We first illustrate the threat model.
Then, we present a high-level description of the vulnerability and the workflow of attack phases.

\subsection{Threat Model}
\begin{figure}[!htb]
\centering 
\includegraphics[width=0.34\textwidth]{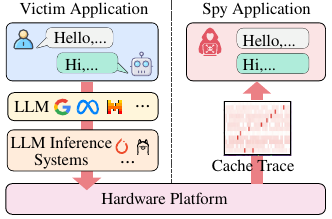}
\caption{The threat model of our eavesdropping attack. Up and down arrows denote information leakage paths. Dashed line in the middle represents OS process isolation.} 
\label{fig:threat_model}
\end{figure}

As shown in \myreffig{threat_model}, we investigate the feasibility of the eavesdropping attack in real-world scenarios where a victim is using a locally deployed LLM.
In this scenario, the operating system isolates the adversary and victim processes.
We assume that the unprivileged adversary cannot compromise the victim's LLM inference system.

\definecolor{info-arrow-color}{RGB}{235, 128, 133}

\myparhead{Adversary} We assume that a spy application made by the adversary can be installed and executed on the victim machine.
Adversaries can create a benign application within popular categories and implant malicious code.
Since the spy application does not tamper with the software libraries used by LLM inference, and only involves legitimate system operations, it can be safely published to the public (e.g., Microsoft Store).
Once launched, the spy application will stealthily eavesdrop on the LLM in the background but will not interact with the victim LLM.
The attack does not require any special privileges, and the spy's virtual address space is isolated from the victim process by the OS.
Consequently, the victim will be unaware of the attack.

\myparhead{Adversary's Capability} 
The adversary cannot trigger requests to the victim LLM and also cannot tamper with the LLM software.
However, aligned with the standard flush+reload attack model, we make the following assumptions:
(\romannumeral1) The adversary can execute malicious code on the victim machine.
(\romannumeral2) The malicious code can open the model file in read-only mode and call \mytechno{mmap}.
(\romannumeral3) If \mytechno{mmap} is unavailable, the adversary can leverage page deduplication to access shared memory.
(\romannumeral4) The adversary can execute the cache line flushing instruction \mytechno{clflush}.

Additionally, the adversary does \emph{not} need to know the model architecture and source code or binary of the LLM inference framework, but he/she has access to the publicly available information to compute the address offset of embedding table elements in the model files.
The rules for deriving the offset are not secret for mainstream LLMs~\cite{safetensors,pytorch-memory-format,pickle-protocol}.

\myparhead{Victim} The victim interacts with a locally deployed LLM for various purposes, such as seeking personal advice and writing sensitive emails.
We assume that the token embedding operation in the LLM inference is offloaded to CPU while other layers of the models can be sent to any device.
This assumption is realistic because it is the default behavior of the mainstream locally-deployed LLM inference frameworks, mainly due to the optimum cost-effectiveness of CPU operations in local deployment scenarios supported by Intel~\cite{intelLlama2}, Meta~\cite{llamaevery}, Numenta~\cite{hotchips-supercharged-ai-cpu}, and the community~\cite{huggingface-cpu-infer, llama.cpp}.

\begin{figure*}[!t]
\centering
\includegraphics[width=0.95\textwidth]{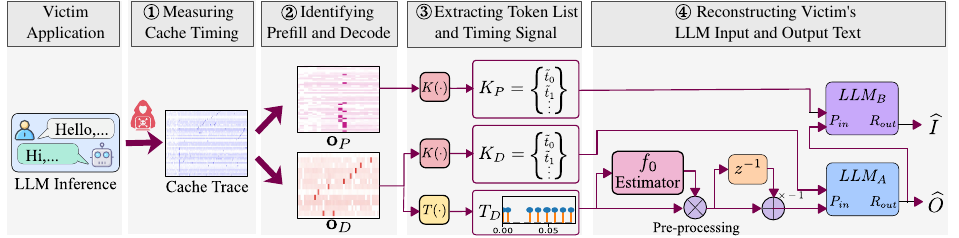}
\caption{Workflow of our eavesdropping attack.}
\label{fig:design_overview}
\end{figure*}

\subsection{The Vulnerability}
\label{sec:intuition}
By studying how LLM inference generally works, we reveal novel side-channel leakage sources that unveil token \emph{value} and \emph{position}, enabling reconstruction of the original text.

\myparhead{Token Value Leakage}
LLMs depend on token embedding to convert token sequences into semantic representations that the model can process.
The token embedding is essentially a linear projection that maps one-hot-encoded token lists into its dense representation. 
Suppose that the token list is encoded as the one-hot matrix $\mathbf{x}$, where $\mathbf{x}_i=[0\cdots 1_{(t_i)}\cdots 0]^T$ and $t_i$ is the i-th token index.
Then, the token embedding is formulated as $\mathbf{E} = \mathbf{W}\mathbf{x}$.

Due to the sparsity of one-hot encoding, it is common to simplify the matrix multiplication into the table lookup $E_i = \textbf{W}[t_i]$, i.e., retrieving the $t_i$-th row of $\mathbf{W}$.
Since only the requested row vectors are loaded and cached on the chip, adversaries can infer the token index $t_i$ by monitoring which row of embedding table $\mathbf{W}$ has been recently accessed by the victim.

Moreover, due to the autoregressive characteristic, all the newly produced tokens are \emph{fed back} into the model, where each new token is sent to the token embedding operation.
Therefore, the token embedding operation leaks both the victim's input and output tokens.

Without any privilege, the spy process cannot directly monitor the memory access of the victim process.
Fortunately, cache side-channel attacks can be exploited to indirectly infer the victim's memory access.
Once a cache hit of the $j$-th row of $\mathbf{W}$ is observed, it can be inferred that the embedding operation has processed the token with the index of $j$.
Moreover, the token embedding operation is typically offloaded into the CPU, due to the optimum cost-effectiveness of CPU computing.
In contrast, other layers of the model can be offloaded to GPUs.
Hence, the adversary can mount cache side-channel attacks on CPU only to cover scenarios where the victim uses GPU to accelerate the model inference.

\myparhead{Token Order Leakage} The above leakage unveils values of input and output tokens, but their \emph{positions} in the original sequence remains unknown.
We seek another leakage source in the temporal domain to recover the token order.
Specifically, LLM inference generally follows the autoregressive paradigm, which consists of a prefill phase and a series of decode phases.
These phases are inherently serialized.
The decode phases will not begin until the prefill is complete.
Similarly, the $i+1$-th decode phase will wait for the completion of the $i$-th token.
Therefore, the time points of these distinct phases are distinguishable in the time dimension, leaking: (1) the boundary between input and output tokens, and (2) the order of output tokens.

However, it is non-trivial to implement the attack by exploiting the above leakages, given the following unique challenges:

\myparhead{Challenge 1 (C1)} \emph{Noise in the side channels introduces errors.}
We deduce the token values by monitoring the cache access of each row of $\mathbf{W}$.
However, the cache side-channel presents noise, i.e., false positives and false negatives, which leads to randomly-valued and missing tokens.

\myparhead{Challenge 2 (C2)} \emph{From the side-channel observer perspective, the order of input tokens is scrambled.}
During model inference, input tokens of the LLM are batched and computed in parallel.
Therefore, the token embedding operation for each input token is randomly interleaved or overlapping in the time axis, prohibiting adversaries from restoring the token positions via the timing of token embedding operations.

\subsection{Attack Workflow}
In this work, we address \textbf{C1} by fine-tuning $LLM_A$ on a synthetic dataset to capture the cross-modality features, namely timing signal and token text, with the aim of reconstructing the original text from the noisy cache trace.
We tackle \textbf{C2} by fine-tuning $LLM_B$ on the synthetic dataset to exploit the contextual dependence between model output and input, and restore the position of input tokens.
The workflow of our attack consists of the following phases, shown in \myreffig{design_overview}.

\begin{enumerate}
\item Execute the spy process that co-locates with the victim and collects the cache trace $\mathbf{o}$ during LLM inference.
\item Identify cache trace segments $\mathbf{o}_P$ and $\mathbf{o}_D$ that correlate with the prefill and decode phases of the victim.
\item Map the cache trace to the ordered token list $K_D$ and the timing signal $T_D$ for the decode phase of the victim.
Additionally, derive the unordered token list $K_P$ for the prefill phases.
\item Reconstruct the victim's model output by fusing $K_D$ and $T_D$ and infer the text via $\widehat{O}=LLM_A(K_D, T_D)$.
\item Reconstruct the victim's model input by fusing $\widehat{O}$ and $K_P$, and deduce the text by $\widehat{I}=LLM_B(\widehat{O}, K_P)$.
\end{enumerate}

\section{Attack Implementation}
In this section, we elaborate on the implementation of the attack phases presented in the workflow.
\subsection{Measuring Cache Timing}
\label{sec:cache_probing}

To exploit the leakage source described in \myrefsec{intuition}, the adversary should infer the victim's cache accesses to $|V|$ rows in the embedding table $\mathbf{W}$.
These accesses can be deduced from the cache trace that is captured by shared-memory-based cache side-channel attacks.
We use this kind of attack because it offers spatial granularity of cache line.
This fine spatial granularity enables the differentiation of distinct rows in $\mathbf{W}$.
In this work, we use flush+reload~\cite{flush+reload}, but we expect that the attack can also be implemented via other shared-memory cache attacks, such as flush+flush~\cite{flush+flush} and evict+reload~\cite{usenix15/cache-template}.

Before mounting the attack, the adversary should compute a total of $|V|$ target addresses using publicly available information about the model file format~\cite{safetensors,pytorch-memory-format,pickle-protocol}.
Suppose that the start address of the shared memory region is $p_1$, and the address offset of $\mathbf{W}$ is $p_2$.
For the $i$-th row of $\mathbf{W}$ ($0<i<|V|$),
if the $\mathbf{W}$ is in row-major order, the target address $A_i$ can be selected in the range $p_1+p_2+iDb \le A_i<p_1+p_2+(i+1)Db$, where $D$ is the dimension of the embedding vector, and $b$ is the size of the vector element.
Similarly, if the $\mathbf{W}$ is in column major order, we select the address $A_i\in\{p|p=p_1+p_2+(i + j|V|)b,0\le j< D\}$.
For each $i$, only one deterministic address within the ranges or sets is chosen.
However, we retain the ranges and sets and will resolve $A_i$ in the next phase.

\myparhead{Evading Hardware Prefetchers}
Interestingly, we reveal that the standard flush+reload implementation failed to maintain the required precision on Intel's recent microarchitecture, namely Raptor Lake.
After implementing state-of-the-art adversarial strategies, the standard flush+reload implementation~\cite{masikt} achieves near-zero precision in probing $|V|=32768$ addresses (the vocabulary size of Llama2), as the average false positive count reaches 5,612,394 per second, indicating high noise from prefetcher activities.

The root cause of this failure is that Intel has employed Array-of-Pointers (AoP) prefetchers in hardware~\cite{gofetch24}, which unfortunately degrades the flush+reload attacks.
Specifically, the attack program depends on a pointer array to store the shuffled target addresses, which is required to overcome streaming and spatial prefetchers~\cite{masikt}.
When the program visits the array, the AoP prefetcher automatically dereferences the memory pointers in the pointer array, leading to a cache hit for almost every target address.

To overcome the AoP prefetchers, we calculate the minimum value of all target addresses as the offset, subtract this offset from each address, and store the subtracted results in an array.
When each target address is required in the attack, the original address is restored by adding the offset to the corresponding value in the array.
This ensures that the array no longer contains valid pointers, thereby preventing prefetch triggering by the AoP.

For other types of prefetcher, we follow previous works, i.e., we choose the target addresses $A_i$ such that no adjacent pages are probed~\cite{usenix15/cache-template}, and randomize the order of probing sequence~\cite{cache-occupancy19}.

\myparhead{Allocate Shared Memory} To achieve the shared memory relied on the cache attacks, we employ two orthogonal strategies:
(\romannumeral1) We note that the memory content of $\mathbf{W}$ is loaded from model files on the disk.
Currently deployed LLM frameworks typically adopt the zero-copy loading technique to optimize the loading performance, eliminating extra data moving via virtual memory mapping and demand paging.
Inspired by this observation, the spy process can share memory with the victim by simply calling \mytechno{mmap} on the model file.
Once the victim process maps to the same file, the adversary will automatically share the physical frames of the file with the victim due to the page cache.
In this case, page duplication is unnecessary.
(\romannumeral2) In rare cases where the victim does not support zero-copy model loading, we utilize page deduplication~\cite{arcangeli2009increasing,singleton} provided by OS to obtain the shared memory.

A detailed pseudocode for the trace acquisition process using standard flush+reload with conventional multithreaded partitioning is provided in the \myrefapp{probing-pseudo-code}.

\begin{figure}[!htb]
\centering 
\includegraphics[width=0.45\textwidth]{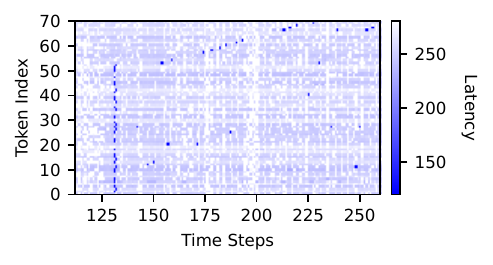}
\caption{An example of cache trace. A deeper color indicates a higher probability that the token embedding operation accesses the token. For clarity, tokens are sequentially re-indexed.} 
\label{fig:cache_trace}
\end{figure}

\myparhead{Example of Cache Trace} The obtained cache trace $\mathbf{o}$ is a $L\times |V|$ matrix, where each row denotes the memory accessing latency of $|V|$ target addresses at $L$ time points, which is shown in \myreffig{cache_trace}.
Additionally, the physical timestamp of each time point is captured by \mytechno{rdtsc} instruction and recorded in the vector $\mathbf{t}$ to facilitate subsequent attack phases.

\subsection{Identifying Prefill and Decode Phase}

According to our threat model, the spy process does not need to interact with the victim process, i.e., it neither controls the start of the victim process, nor conducts inter-process communication with the victim.
Like preceding studies, this setting belongs to asynchronous attack~\cite{cache-atk-aes-sp11}.
Therefore, the adversary needs to identify the start time of LLM inference phases, particularly the boundary of prefill and decode phases.

We design a pattern-matching algorithm to find the cache trace segment $\mathbf{o}_P$ and $\mathbf{o}_D$ that correlates with prefill and decode phases, respectively.
Our algorithm exploits the behavior of autoregressive generation.
Particularly, prefill phase batches input tokens as a whole and compute token embedding in parallel.
Therefore, the prefill phase causes abnormally \emph{dense} cache hit events that cluster around a short period (e.g., around time step 130 in \myreffig{cache_trace}).
Meanwhile, decode phases produce output tokens \emph{serially}, correlating with cache hit events much \emph{sparser} than the prefill phase in the time axis (e.g., one cache hit per time point).

We match $\mathbf{o}_P$ and $\mathbf{o}_D$ in the $\mathbf{o}$ via time interval between each two consecutive cache hit events, i.e., $\mathbf{t}_i - \mathbf{t}_{i-1}, 0<i<L$.
The start point of $\mathbf{o}_P$ is matched if no less than $K$ consecutive events after the point have time intervals that are all less than a threshold $\alpha_1$.
Meanwhile, the start point of $\mathbf{o}_D$ is matched if at least one subsequent event has a time interval greater than $\alpha_1$.
The parameter $K$ is the minimum allowed number of input tokens ($K=4$ in our implementation), and the threshold $\alpha_1$ is no greater than the minimum time of one model forward propagation ($\alpha_1=10^{-3}$ in our settings).

\subsection{Mapping Token Lists and Timing Signal}
\label{sec:extract_tkl_ts}
Having obtained the $\mathbf{o}_P$ and $\mathbf{o}_D$, we now map the cache trace to the token lists and timing signal for further text reconstruction.

As discussed in \myrefsec{intuition}, the cache hit of the target address $A_i$ indicates that the token embedding operation has accessed the $i$-th row of $\mathbf{W}$, and hence implies that the token $i$ has appeared in the model input or output.
We define a cache hit event as the time points where the memory access latency is lower than a predefined threshold $\alpha_2$, following previous cache attacks~\cite{flush+reload}.
The threshold can be obtained by micro-benchmarks for caches~\cite{nanobench}.

\myparhead{Token Lists} The mapping of token lists consists of two steps.
First, we compute $K(\mathbf{o}_D)=[k_1,k_2,\cdots]$, $k_i=\{\mathcal{T}^{-1}(j)|{\mathbf{o}_D}_{ij}<\alpha_2, 0\le j<|V|\}$, where $\alpha_2$ is the cache hit threshold, $\mathcal{T}^{-1}$ is the de-tokenizer that converts the token index into the token text.
Second, we remove all the empty sets in the $K(\mathbf{o}_D)$ to derive the resulting token list $K_D$.
Similarly, we can derive the unordered token list $K_P=[\mathcal{T}^{-1}(j)|{\mathbf{o}_P}_{ij}<\alpha_2,0\le i<L, 0\le j<|V|]$, where $L$ is the length of the cache trace.

\myparhead{Timing Signal}
We extract the timing signal by deriving cache hit events' timestamps corresponding to each token.
Formally, $T_D=[\mathbf{t}_i|K_i(\mathbf{o}_D)\notin\emptyset,0\le i<L]$.

\subsection{Reconstructing Model Output}
\label{sec:recon_llm_out}
As mentioned in \textbf{C1} (\myrefsec{intuition}), the cache side-channel contains noise, specifically false positive and false negative noise, which induces errors in the token lists and complicates the attack.
A \emph{false positive noise} indicates an observed cache hit event that is not attributed to the victim's token embedding operation.
According to our token list mapping approach, false positives induce randomly inserted tokens in the resulting $K_D$ or $K_P$ since the victim has not processed the token at all.
Meanwhile, a \emph{false negative noise} implies that the cache hit is not successfully observed.
Therefore, false negative noise causes missing tokens in the results.

To identify the noise, we scrutinize the timing signal of the cache trace.
Intuitively, the autoregressive token generation can be approximated as \emph{periodic} events, since the forward propagation of LLMs is data-flow-oriented and uses regular pipelines, which differs from control-flow-oriented programs.
To validate the periodicity of the autoregressive generation, we study the timing signal in the frequency domain using Fourier transformation.
Mainly, we utilize the Power Spectral Density (PSD), which is a measure used in signal processing to describe how the power (or ``strength'') of a signal is distributed across different frequencies.
If the signal exhibits periodicity, the PSD will show a peak.

\begin{figure}[!htb]
\centering 
\includegraphics[width=0.45\textwidth]{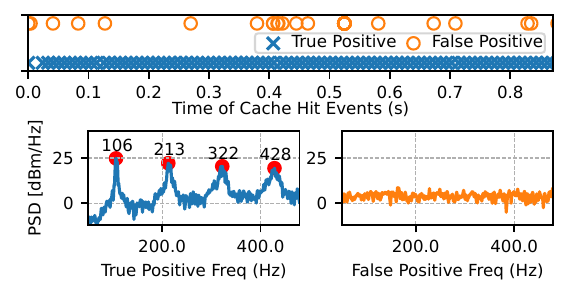}
\caption{PSD of the timing signal derived from the cache trace $\mathbf{o}_D$. The signal is windowed by the Hann function.} 
\label{fig:example_psd}
\end{figure}

We model the \emph{timing signal} of cache trace $\mathbf{o}_D$ as a Dirac impulse train
$T'_D(t) = \sum_{k=1}^{|T_D|} \delta(t - {T_D}_k)$.
Then, we can derive the Fourier transform of the timing signal using the sifting property of Dirac functions:
\begin{displaymath}
F(\omega;T_D) = \int_{-\infty}^{\infty} T'_D(t) e^{-j\omega t} dt = \sum_{k=1}^{|T_D|} e^{-j\omega {T_D}_k}
\end{displaymath}
\noindent which transforms the timing signal into its frequency domain, and further enables the estimation of PSD.
\myreffig{example_psd} shows the PSD estimation of timing signal corresponding to the decode phases of LLM inference.

The top of \myreffig{example_psd} presents the timing signal collected from the llama.cpp that runs Phi-3.5-mini model on an Intel 13900K platform with NVIDIA RTX3060 GPUs.
The average time per output token (TPOT) is about 10 microseconds.
Thereby, we expect a peak at the frequency of $100 \mathrm{Hz}$ in the PSD.

The bottom of \myreffig{example_psd} shows the PSD of the cache traces respective to the true positives and false positives, from the left to the right, respectively.
We clearly observe the expected peak at the base frequency of about $f_0=100\mathrm{Hz}$ and harmonic frequencies at $kf_0, k=2,3,...$ in the PSD of the true positive trace.
Therefore, we argue that the true positives of the investigated decoding phases exhibit strong periodicity.
Meanwhile, in the PSD of false positive trace, the frequency components are relatively evenly distributed across the entire frequency range, implying that false positives are close to white noise.

The PSD offers an effective method for extracting periodic components from the timing signals that contain noise.
To further investigate the remaining noise, we now remove the periodic components by computing the normalized first-order difference of the timing signal, as follows:
\begin{equation}
\label{eqn:norm_td}
\hat{T}_{D_k}=f_0(T_D)\cdot({T_D}_{k}-{T_D}_{k-1})
\end{equation}

\begin{figure}[!htb]
\centering 
\includegraphics[width=0.48\textwidth]{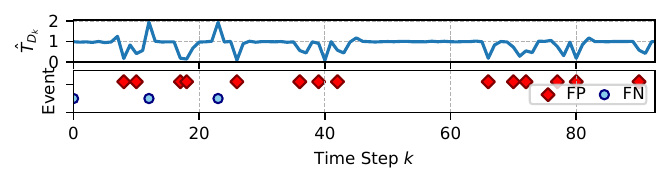}
\caption{An example of the pre-processed timing signal. False positive (FP) noise likely aligns with the valley. False negative (FN) noise likely aligns with the peak.}
\label{fig:example_ntdf}
\end{figure}

Leveraging the PSD, we normalize the first-order difference by multiplying it with the extracted fundamental frequency $(f_0)$.
This normalization renders the results that were previously susceptible to hardware-specific TPOT relatively hardware invariant.
The base frequency is extracted from the PSD using Sawtooth Waveform Inspired Pitch Estimator Prime (SWIPE') algorithm~\cite{SWIPE}. 
The resulting signal is shown in \myreffig{example_ntdf}.

We can find that false positive noise correlates with the peaks in the waveform of $\hat{T}_D$.
The underlying reason is that true positives are periodic, i.e., are highly likely to occur at regular time points near $kT+\varphi$ (where $\varphi$ represents the initial phase).
Assuming that we remove the $i$-th true positive to create a false positive, then a gap would appear between $t_1=(i-1)T$ and $t_2=(i+1)T$, and therefore the first-order difference would rise to $t_2-t_1=2T$.
If such a blank is detected, we can eliminate the corresponding false positive by first predicting its missing token using the context of the preceding and subsequent tokens, then inserting the predicted token to $K_D$.
The token prediction aligns with fill-in-the-blank NLP tasks, which \emph{interpolates} the token sequence.

Similarly, we can find that most false negatives correlate with the valleys of $\hat{T}_D$, because true positives are highly likely to occur at the periodic time points, while false positives are relatively uniformly distributed at the time axis.
A false positive occurring between $t_1=kT+\varphi$ and $t_2=(k+1)T+\varphi$ will result in a first-order difference $t_2-t_1<T$.
If a false positive is detected, it should be removed from the $K_D$.

The analysis implies that, depending on the waveform of $\hat{T}_D$, adversaries must choose one of three actions at each time point: fill a blank, remove a false positive, or leave the token unchanged.
We found that this problem can be uniformly formulated as a sequence-to-sequence task.
Given the timing signal $\hat{T}_D$ and the token list $K_D$, the resulting text is:
\begin{displaymath}
\widehat{O}\sim P(O|\hat{T}_{D_1},{K_D}_1, \hat{T}_{D_2},{K_D}_2,\cdots)
\end{displaymath}
\noindent where $P$ is implemented by the $LLM_A$ fine-tuned to capture the timing patterns of $\hat{T}_D$ and remove noise.

\myparhead{Training Set Synthesis}
We need to synthesize data for the training model $P$ because the adversary cannot access the actual victim models or systems, thereby not being able to profile them to collect  the training data.
Fortunately, the timing patterns revealed by the PSD enable the automatic synthesis of training samples.

\begin{algorithm}[htb]
\caption{Dataset Synthesis}
\label{algo:dataset_synth}
\LinesNumbered
\DontPrintSemicolon
\KwIn{The probability of noise $p$, the standard deviation of duration $\sigma$, the corpus dataset $C$, and the size of LLM vocabulary $|V|$.}
\KwOut{The synthesized training dataset $T$}
\tcp{$U$ denotes uniform distribution, $\mathcal{N}$ is normal distribution. $s(\cdot)$ is a sample.}
\For{$c \in C$} {
	\textbf{Init} $L \gets \phi$ \tcp{Generated cache trace}
	\textbf{Init} $ctime \gets 0$ \tcp{Current timestamp}
	\textbf{Init} $mtime \gets |c| + 1$ \tcp{Maximum length}	

	\For{$token \in c$} {
		\If{$s(U[0,1]) \ge p$} { 
			$L \gets L \cup \{(ctime,token)\}$ 
		}
		\If{$s(U[0,1]) < p$} {
			$L \gets L \cup \{(s(U[0,mtime]),s(U[0,|V|-1])\} $
		}
		$ctime\gets ctime + s(\mathcal{N}(1,\sigma^2))$ \tcp{Simulate the periodicity of decode phases}
	}
	
	$T\gets T \cup \{(L[0], L[1], c)\}$\tcp{Get a ($T_D$,$K_D$,$O$)}
}

\normalsize
\end{algorithm}

The training set synthesis consists of three steps.
First, we collect LLM input and output text pairs $(I,O)$ to form the textual corpus.
We query general-purpose LLMs with various prompts from public datasets (like UltraChat) as input $I$ to obtain the model output $O$.
Second, we leverage the periodicity revealed by the PSD (\myreffig{example_psd}) to generate the simulated timing signal $T_D$.
As shown in \myrefalgo{dataset_synth}, we generate the timestamps of token generation by accumulating Gaussian noise that simulates the minor fluctuations of TPOT.
Meanwhile, we randomly add false positives and false negatives with probability $p$ to simulate the noise, and generate $K_D$
Finally, after generating the training samples $(T_D,K_D,O)$, we encode the samples into LLM training pairs $(P,R)$ that consist of prompts and ground-truth responses.
We encode the numerical timing signal $T_D$ as textual prompts, similar to the previous work on time series~\cite{NEURIPS2023_3eb7ca52,nips24-promptcast}.
For example, we encode the $T_D=\{1.3,1.2\}$ and $K_D=\{'Hello','World'\}$ into the following textual prompt:
\begin{center}
\{(1.3,'Hello'),(1.2,'World')\} $\rightarrow$ '1 3:Hello\mytechno{<s>}1 2:World\mytechno{<s>}'
\end{center}

\myparhead{LLM Fine-tuning}
Having obtained the synthesized training set, we fine-tune the base LLM (e.g., Llama) to obtain the $LLM_A$.
We utilize LoRA to fine-tune the LLM, which allows us to retain most of the knowledge from the pre-trained model while mitigating overfitting.
Finally, during the attack, the adversary can obtain the reconstructed model output of the victim via $\widehat{O}=LLM_A(T_D,K_D)$, as shown in \myreffig{design_overview}.

\subsection{Reconstructing Model Input}
\label{sec:recon_llm_in}
Unlike the output text, reconstructing the input text is more challenging.
As mentioned in \textbf{C2} (\myrefsec{intuition}), the parallel computing of token embedding operation results in \emph{unordered} input tokens in the side-channel attacking results.

The model of unordered tokens aligns with the bag-of-words language model.
A bag-of-words model discards the sequential order and assumes that tokens are independent of one another.
Without the token order, the bag-of-words model can still reveal latent semantic structures, and is widely applied in the topic models.
For instance, Latent Dirichlet Allocation (LDA) automatically identifies topic or theme words from a collection of documents, aiding in understanding the main content of textual data~\cite{topic-modeling-lda}.

However, the loss of token order can often impact the linguistic structure and precise semantics of sentences.
For instance, the sentences "A chased B" and "B chased A" have the same token set but different semantics.
We expect to reconstruct the full text of the model input instead of a series of discrete words.

Fortunately, the LLM input generally correlates with its output in the same context.
For example, LLMs tend to repeat the question before proceeding~\cite{remote-kla-2024}.
We formulate the problem of reconstructing model input as a sequence-to-sequence model that restores the positions of each input token.
The model is as $\widehat{I} \sim P(I|K_P,\widehat{O})$,
where $K_P$ is the unordered set of input tokens, and $\widehat{O}$ is the reconstructed model output, $P$ is implemented by the $LLM_B$ fine-tuned on a synthetic training set.

We encode the token list $K_P$ as a text sequence for the LLM prompt delimited by \mytechno{<s>} token.
Then, we insert the reconstructed model output $\widehat{O}$ at the beginning of the encoded $K_P$, providing the context for the model to infer the token order.

\myparhead{Training Set Synthesis} Similar to the strategy of dataset synthesis in \myrefsec{recon_llm_out}, we randomly shuffle token positions of the prompt $I$ in the corpus dataset to obtain $K'_P$, which simulates the characteristic of cache trace obtained from the prefill phase.
Then, we feed the prompt into a general LLM as its input to obtain the corresponding output text $O'$, and add the training sample $(K'_P,O',I)$ to the resulting dataset, where $I$ is the ground-truth of the input reconstruction.

\myparhead{Fine-tune LLM}
We fine-tune the pre-trained LLM to obtain the $LLM_B$, in the similar process described in \myrefsec{recon_llm_out}.
Finally, during the attack, the adversary can obtain the reconstructed model input of the victim via $\widehat{I}=LLM_B(K_P,\widehat{O})$, as shown in \myreffig{design_overview}.

\section{Experiment Setup}
\label{sec:exp-setup}

In this section, we introduce the experiment setup of the evaluations for our eavesdropping attack framework.

\myparhead{Environment}
We tested the attack on the machine equipped with an Intel(R) 13-Gen 13900K 5.8GHz CPU (with the latest BIOS microcode: 0x129), an NVIDIA 3060 GPU (with 12GiB GPU memory), and 32GiB of dual-channel DDR4-3200 memories on a Maxsun(R) H610ITX baseboard.
In \myrefsec{hardware_eval}, CPUs (except for the 13900K) were tested on a Gigabyte(R) H610M K DDR4 motherboard equipped with an NVIDIA 3060 GPU and 16GiB of single-channel memory.
These machines runs Ubuntu 22.04 (with Linux 5.18.0-38-generic kernel) OS, NVIDIA GPU driver v525.89.02, and CUDA v12.0.

\myparhead{Dataset Construction}
We synthesize the dataset for fine-tuning $LLM_A$ and $LLM_B$ via processes described in \myrefsec{recon_llm_out} and \myrefsec{recon_llm_in}.
Specifically, we sample prompt text $I$ from UltraChat~\cite{ultrachat}, NQ-Open~\cite{kwiatkowski2019natural} (belonging to Natural Question), SIQA~\cite{sap-etal-2019-social}, SQuAD2~\cite{rajpurkar-etal-2018-know}, and ChatGPT-Roles~\cite{ChatGPT-Roles} datasets for our purpose.
The synthesized dataset was randomly partitioned into training, validation, and test sets at ratios of 60\%, 20\%, and 20\%, respectively.
Details of dataset synthesis settings and data cleaning can be found in \myrefapp{dataset-construction}.

\myparhead{Model Fine-tuning}
The fine-tuning is conducted on an AutoDL cloud server with NVIDIA H800 computing card, 300GB CPU memory, and AMD EPYC 9K84 CPU.
Hyperparameters of the fine-tuning can be found in \myrefapp{finetune-hyperparams}.

\myparhead{Metrics} We evaluate the fidelity of the reconstructed model input and output using metrics of different granularities:
In \emph{character-level}, we used Levenshtein Similarity (LS = 1 - Edit Distance).
In \emph{token-level}, we used ROUGE, a measure of n-gram similarity between the reconstructed text and the ground-truth.
Specifically, we used R1 and RL metrics, which evaluate F1-score via n-grams and the longest sub-sequence.
In \emph{semantic-level}, we employed cosine similarity ($\phi$) of sentence embedding to evaluate how effectively the semantics of the text have been reconstructed.
We employed the state-of-the-art embedding model AngleIE-Llama-7B~\cite{AngleIE} to compute the sentence embedding of both the ground-truth text and the reconstructed text, then compute their normalized dot product to obtain the cosine similarity $-1\le \phi \le 1$, where $\phi=1$ indicates a perfect match.

\begin{figure}[!htb]
\centering 
\includegraphics[width=0.45\textwidth]{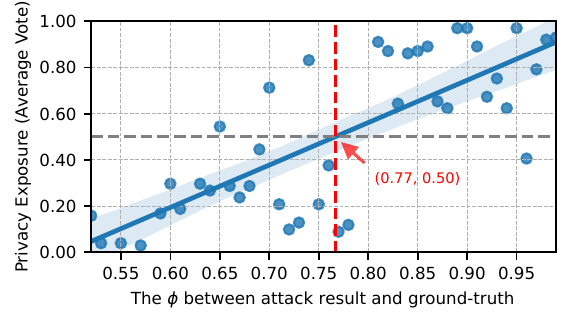}
\caption{The relationship between human-evaluated privacy exposure and cosine similarity, with a Pearson Correlation of 0.771.}
\label{fig:survey}
\end{figure}

To objectively determine which value of $\phi$ indicates a successful attack where the semantics of the text have been leaked, we conducted a survey study inspired by previous works~\cite{remote-kla-2024}.
Using test results, we uniformly sampled 50 sentences across the range $\phi=[0.5,1.0]$, and recruited 100 random participants on Prolific to vote on whether the attack results accurately capture the privacy contents of the ground-truth sentence.
\myreffig{survey} plots the relationship between the human-evaluated privacy exposure and the cosine similarity.
This linear relationship indicates that a majority of participants believe privacy information is accurately captured when $\phi$ is above 0.77. Thus, we define the attack success rate (ASR) as the proportion of testing samples where $\phi > 0.77$.

\section{Evaluation}
In this section, we evaluate the threat of our LLM eavesdropping attack and aim to answer the following Research Questions (RQs):
\begin{itemize}
\item RQ1 [Attack Performance] What is the attack performance against different types of victim LLMs in practical systems?
\item RQ2 [Parameter Analysis] How do different hyperparameters affect the attack performance?
\item RQ3 [Ablation Study] What is the contribution of key components  to the attack performance?
\item RQ4 [Framework Evaluation] Which LLM frameworks are affected by the side-channel vulnerability?
\item RQ5 [Hardware Evaluation] What is the applicability of the attack to other machines?
\item RQ6 [Attack Examples] How can concrete examples be used to understand the privacy leakage?
\end{itemize}

\subsection{RQ1: Attack Performance in Practical Systems}

\begin{table*}[!htb]
	\centering 
	\caption{Attack performance of each victim LLM, using GPT-4o-mini-2024-07-18 as the base model of $LLM_A$ and $LLM_B$. $N_I$ and $N_O$ denotes the average number of input and output tokens, respectively. An \underline{underline} represents the minimum value of each metric, while the maximum value is represented by \textbf{bold} text.}
	\label{tab:atk_result}
	\small
	\setlength{\tabcolsep}{3.2pt}
		\begin{tabular}{c|c|ccccccccccccc}
			\toprule
			\multirow{2}*{Victim LLM} & \multirow{2}*{Dataset} &
			\multicolumn{6}{c}{Output Reconstruction} & & \multicolumn{6}{c}{Input Reconstruction}  \\ \cline{3-8} \cline{10-15} 
			& & $N_{O}$ &  R1 (\%)  & RL (\%) & LS (\%) & $\phi$ (\%) & ASR (\%) &  & $N_I$ &   R1 (\%)  & RL (\%) & LS (\%) & $\phi$ (\%) & ASR (\%) \\

\midrule
\multirow{4}*{\begin{tabular}{@{}c@{}}Google \\Gemma2-9B\\~\cite{gemma}\end{tabular}} & UltraChat & 243& 98.2& 98.2& 97.0& 99.6& 99.8  & & 20& 93.5& 90.2& 87.4& 99.2&  \textbf{100.0}        \\
& NQ-Open & 79& 95.9& 95.9& 94.3& 98.7& 99.3    & & 13& 94.6& 93.0& 91.3& 99.0&  \textbf{100.0}        \\
& SIQA & 193& 96.4& 96.4& 94.2& 98.8& 99.1      & & 31& 86.6& 79.2& 74.8& 96.9&  \textbf{100.0}        \\
& SQuAD2 & 55& 91.5& 91.5& 89.8& 98.2&  \textbf{100.0}  & & 183& 57.1& 47.7& 34.4& 94.9&  \textbf{100.0}       \\
& ChatGPT-Roles & 222& 98.7& 98.7& 98.0& 99.6&  \textbf{100.0}  & & 48& 85.4& 79.7& 70.6& 99.1&  \textbf{100.0}        \\
\midrule
\multirow{4}*{\begin{tabular}{@{}c@{}}Meta \\Llama-3.1-8B\\~\cite{llama}\end{tabular}} & UltraChat & 253& 99.0& 99.0& 98.9& 99.2& 99.3 & & 19& 94.5& 91.9& 89.5& 99.2&  \textbf{100.0}        \\
& NQ-Open & 162& 97.4& 97.4& 96.9& 98.1& 98.0   & & 12&  \textbf{94.8}&  \textbf{93.4}&  \textbf{91.4}& 99.0&  \textbf{100.0}  \\
& SIQA & 64& 98.1& 98.1& 97.6& 98.9& 99.1       & & 30& 86.1& 78.5& 73.6& 96.6& 99.7  \\
& SQuAD2 & 20&  \underline{90.1}&  \underline{90.1}& 90.4&  \underline{96.7}&  \underline{96.4}        & & 180& 55.8& 46.4& 33.2& 94.3&  \textbf{100.0}        \\
& ChatGPT-Roles & 215&  \textbf{99.5}&  \textbf{99.5}&  \textbf{99.6}&  \textbf{99.8}&  \textbf{100.0} & & 48& 86.3& 80.7& 72.3& 99.0&  \textbf{100.0} \\
\midrule
\multirow{4}*{\begin{tabular}{@{}c@{}}TII \\Falcon3-10B\\~\cite{falcon3}\end{tabular}} & UltraChat & 175& 98.4& 98.4& 97.3& 99.6& 99.6 & & 20&  \textbf{94.8}& 92.1& 90.2&  \textbf{99.3}&  \textbf{100.0}    \\
& NQ-Open & 109& 98.2& 98.1& 97.7& 99.7& 99.9   & & 13& 94.3& 92.6& 91.1& 99.0&  \textbf{100.0}        \\
& SIQA & 140& 98.9& 98.9& 97.9& 99.7&  \textbf{100.0}   & & 31& 86.2& 78.6& 75.5& 96.7&  \textbf{100.0}        \\
& SQuAD2 & 62& 90.6& 90.6& 93.2& 98.0&  \underline{96.4}        & & 185& 54.6& 44.9& 33.5& 93.8&  \textbf{100.0}       \\
& ChatGPT-Roles & 67& 98.9& 98.8& 99.3& 99.6&  \textbf{100.0}   & & 48& 86.8& 82.3& 73.9& 99.0&  \textbf{100.0}        \\
\midrule
\multirow{4}*{\begin{tabular}{@{}c@{}} Mistral-7B\\~\cite{mistral}\end{tabular}} & UltraChat & 256& 94.6& 94.6& 91.6& 98.2& 98.7       & & 20& 91.6& 87.7& 84.4& 98.7&  \textbf{100.0}        \\
& NQ-Open & 120& 95.1& 95.1& 94.6& 97.1& 96.8   & & 12& 89.1& 84.0& 80.8& 97.3& 99.8  \\
& SIQA & 65& 98.7& 98.7& 98.2& 99.4& 99.7       & & 32& 85.9& 77.6& 73.7& 96.2&  \textbf{100.0}        \\
& SQuAD2 & 57& 91.4& 91.4& 90.1& 96.9& 98.2     & & 204& 51.3& 43.2& 32.4& 92.7& 98.2 \\
& ChatGPT-Roles & 243& 94.6& 94.6& 91.6& 98.9&  \textbf{100.0}  & & 54& 83.2& 78.4& 69.7& 97.9&  \textbf{100.0}        \\
\midrule
\multirow{4}*{\begin{tabular}{@{}c@{}}Microsoft \\Phi-3.5-mini-3B\\~\cite{phi35}\end{tabular}} & UltraChat & 263& 93.5& 93.5& 88.9& 99.0&  \textbf{100.0}      & & 21& 90.5& 87.2& 84.3& 98.2& 99.6   \\
& NQ-Open & 194& 93.9& 93.9& 90.9& 98.7& 99.3   & & 12& 88.0& 82.9& 79.8& 97.0& 99.8  \\
& SIQA & 253& 92.7& 92.7&  \underline{87.7}& 98.5& 99.4 & & 33& 85.2& 78.5& 75.4& 96.5& 99.7   \\
& SQuAD2 & 137& 93.5& 93.5& 90.6& 97.6& 98.2    & & 209&  \underline{51.0}&  \underline{42.4}&  \underline{32.2}&  \underline{92.1}&  \underline{96.4} \\
& ChatGPT-Roles & 263& 94.6& 94.6& 92.1& 98.8&  \textbf{100.0}  & & 57& 80.6& 75.0& 65.7& 97.6&  \textbf{100.0}        \\
\midrule
\multicolumn{2}{c}{Average} & 165& 96.3& 96.3& 94.8& 98.7& 99.1 & & 24& 89.9& 85.8& 82.7& 98.0& 99.9   \\

			\bottomrule
		\end{tabular}
\end{table*}
In the evaluation of attack performance, we execute the victim and co-located spy process on the victim machine.
The victim process runs a real-world LLM framework (llama.cpp in our settings).
During experiment, we send testing prompts to the victim process, and wait for the model output, acting as a regular user.
Meanwhile, the spy process performs the eavesdropping attack.
The setting parameters of llama.cpp are kept by default, with a batch size of 256 for the prefill.
\myreftab{atk_result} shows the resulting performance of attacking different victim models, using GPT-4o-mini as the base model of $LLM_A$ and $LLM_B$.

\myparhead{[RQ1-1] Reconstruction Accuracy} We first analyze the performance of output reconstruction.
\myreftab{atk_result} shows that the ROUGE-1 and ROUGE-L values of the output reconstruction are all higher than 90.1\%, and the Levenshtein similarity is no less than 87.7\%, which means that our method can restore high-fidelity target output text at both the character and token levels.
We can observe that cosine similarity $\phi$ for all 5 types of victim models is higher than 96.7\%, implying that the semantics of text has been leaked.
The lowest ASR is 96.4\%, indicating that we can successfully infer most of the privacy contents of the tested samples.

We then describe the performance of input reconstruction.
As presented in \myreftab{atk_result}, the average ROUGE-1 and Levenshtein similarity reach 89.9\% and 82.7\%, indicating reasonable token-level and character-level reconstruction accuracy.
Meanwhile, our method achieves high cosine similarity (the worst is 92.1\%) to the ground-truth text.
We can observe that the average ASR achieves 99.9\%, meaning the topics of 99.9\% of tested sentences were leaked.

Nevertheless, we observed a negative correlation between the number of input tokens and the attack performance.
In the worst case (Phi-3.5-mini on SQuAD2), the ROUGE-1 and Levenshtein similarity metrics drop to 51.0\% and 32.2\%.
This can be attributed to the reconstruction challenge, which resolves the results in a search space that increases factorially, i.e., on the order of  $O(N_{I}!)$.
Nonetheless, the worst cosine similarity reaches 92.1\%, since our model can restore the semantics of original sentences using different textual expressions and can still leak the sentences' topic or semantics.

\begin{figure}[!thb]
\centering
\includegraphics[width=0.42\textwidth]{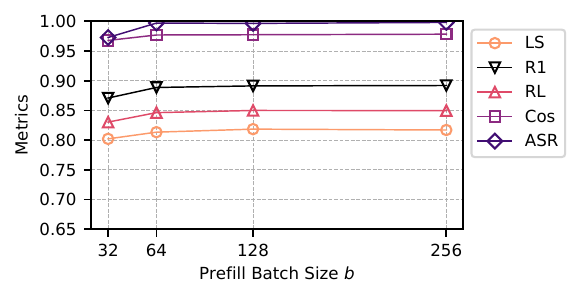}
\caption{The correlation between the prefill batch size $b$ and the attack performance. We use gpt-4o-mini-2024-07-18 as the base model. The tested victims are Gemma2-9B and Phi-3.5-mini-3B.}
\label{fig:data_sens_b}
\end{figure}

\myparhead{[RQ1-2] Different Prefill Bach Sizes}
We investigate the impact of varying prefill batch sizes ($b$), as shown in \myreffig{data_sens_b}.
As $b$ goes up exponentially, the evaluation metrics settle on stable values with a range of only 4\%.
However, when $b$ is small, a slight decline in overall accuracy is noted.
This outcome may be attributed to a shift in the training data distribution.
Specifically, the synthetic training data involves a random permutation of input tokens, which aligns well with real-world scenarios that usually use larger $b$ values.
In contrast, smaller $b$ values necessitate execution through multiple sequential batches when the number of input tokens exceeds $b$.
In such cases, the extracted input tokens $T_P$ display increased sequentiality, deviating from the training data distribution.
Fortunately, empirical results suggest that this deviation does not lead to significant performance degradation.

\subsection{RQ2: Parameter Analysis}
In this research question, we evaluate the impact of hyperparameters in our method.

We have introduced two hyperparameters: $p$ and $\sigma$ involved in the dataset synthesis (\myrefalgo{dataset_synth}) for the output reconstruction.
To evaluate how these hyperparameters affect attack performance, we conducted a series of training sessions on Llama3.1-8B-Instruct using a set of different hyperparameters.
All the evaluations in the following experiments used Gemma2-9B and Phi-3.5-mini as victim models.

\begin{figure}[!htbp]
	\centering
	\subfloat[Output Reconstruct ($\sigma=0.08$).]{%
		\includegraphics[width=0.5\columnwidth]{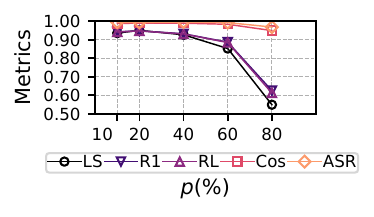}
		\label{fig:param_p_analysis}
	}
	\subfloat[Input Reconstruction ($\sigma=0.08$).]{%
		\includegraphics[width=0.5\columnwidth]{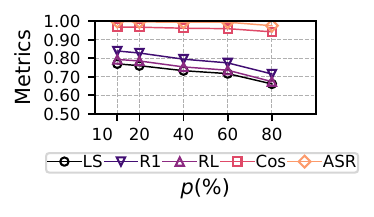}
		\label{fig:param_p_analysis}
	}
	\\
	\subfloat[Output Reconstruct ($p=0.2$).]{
		\includegraphics[width=0.5\columnwidth]{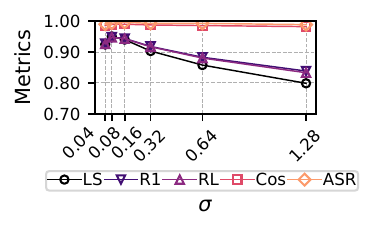}
		\label{fig:param_s_analysis}
	}
	\subfloat[Input Reconstruction ($p=0.2$).]{
		\includegraphics[width=0.5\columnwidth]{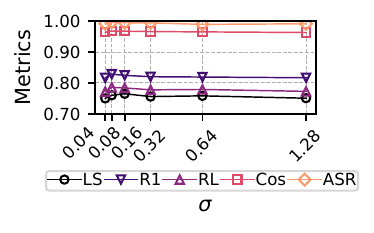}
		\label{fig:param_s_analysis}
	}
	\caption{Correlation between attack performance and the parameter $p$ and $\sigma$. We use Llama3.1-8B-Instruct as the base model of $LLM_A$ and $LLM_B$.}%
	\label{fig:param_p_s_analysis}
\end{figure}

\myparhead{[RQ2-1] Different Values of $p$} We first analyze the sensitivity of the hyperparameter $p$ with $\sigma$ set to a fixed value (0.08 in our settings).
We synthesized several training sets using different $\sigma$.
For each training set, we fine-tuned a pair of $LLM_A$ and $LLM_B$, then evaluated these models on the validation set.

\myreffig{param_p_analysis} shows the evaluation results.
As $p$ increases exponentially, we observe that the overall accuracy remains consistently high when $p$ is set in a wide range (10\% to 40\%).
When $p$ reaches larger values (>50\%), the overall performance begins to decline.
This is likely because more than half of the simulated cache hit events are false positives, introducing additional noise during training and requiring more data for the model to converge.
The findings suggest that in practical scenarios, the parameter $p$ can be safely set within the range of 10\% to 40\%.

\myparhead{[RQ2-2] Different Values of $\sigma$} 
We then analyze the hyperparameter $\sigma$.
For our purpose, $p$ is fixed to 0.1, and we fine-tuned Llama3.1 on a set of $\sigma$ values.
\myreffig{param_s_analysis} shows the testing result on the validation set.

We can observe that the output reconstruction performance peaks at $s=0.08$ and then declines. This occurs because $\sigma$ controls the standard deviation of the decode phase period, normalized to $[0, 1]$ in \myrefeqn{norm_td}.
A smaller $\sigma$ implies greater period certainty; $\sigma=0$ enforces strict periodicity, while $\sigma>0.5$ weakens periodicity and hinders the model's ability to capture desired patterns.
Also, the performance of reconstructing the input is not very sensitive to $\sigma$ because the output reconstruction only serves as an extra reference context.

\myparhead{[RQ2-3] Different Base Models}
To study how model selection impacts the attack performance, we finetuned and compared two base models: open-source Llama-3.1-8B and close-source GPT-4o-mini.

\begin{figure}[!thb]
\centering
\subfloat[Output reconstruction ($LLM_A$).]{%
	\includegraphics[width=0.5\columnwidth]{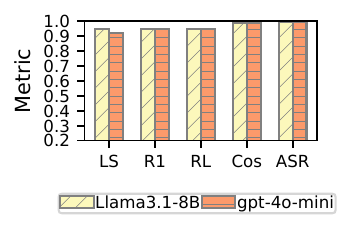}
	\label{fig:model_selection_rr}
}
\hspace{-0.5cm} 
\subfloat[Input reconstruction ($LLM_B$).]{
	\includegraphics[width=0.5\columnwidth]{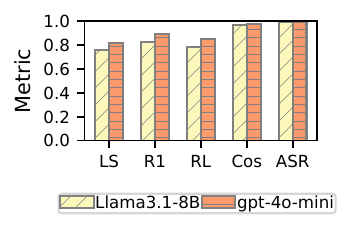}
	\label{fig:model_selection_pr}
}
\caption{Comparative evaluation of different base models.}
\label{fig:model_selection}
\end{figure}

\myreffig{model_selection} shows that our $LLM_A$ and $LLM_B$ fine-tuned on two different base models exhibit relatively consistent attack performance.
We find that specific model architectures decouple the text reconstruction performance.

\subsection{RQ3: Ablation Study}
\label{sec:ablation}
To demonstrate the effectiveness of each component, we conduct the ablation study.
We removed each critical attack phase and re-evaluated the attack performance, to demonstrate the contribution of the proposed components.

\begin{figure}[htb]
\centering
\subfloat[Output reconstruction ($LLM_A$).]{%
	\includegraphics[width=0.5\columnwidth]{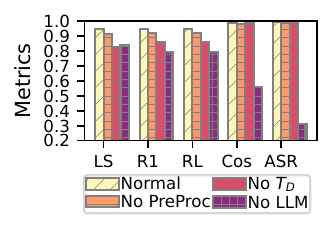}
	\label{fig:ablation_study_rr}
}
\hspace{-0.5cm} 
\subfloat[Input reconstruction ($LLM_B$).]{
	\includegraphics[width=0.5\columnwidth]{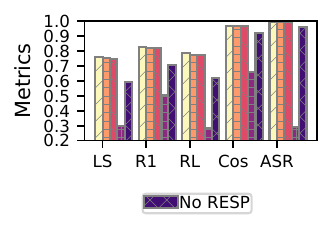}
	\label{fig:ablation_study_pr}
}
\caption{Ablation study using Llama3.1-8B-instruct as the base model of $LLM_A$ and $LLM_B$.}
\end{figure}

\myparhead{[RQ3-1] Ablation Study of Output Reconstruction}
As shown in \myreffig{ablation_study_rr}, after only removing the pre-processing phase (\myrefeqn{norm_td}) of timing signal $T_D$, the Levenshtein similarity dropped -3.0\% and -1.0\% for the output and input reconstruction, respectively.
After removing timing signal $T_D$ (do not include $T_D$ in the encoded prompts) in the output reconstruction, the Levenshtein similarity dropped -12.7\% (output) and -1.2\% (input).
These results show that fusing the token list with the timing signal leads to better performance.
Additionally, we completely removed the $LLM_A$ (assembling the resulting text from the $K_D$ only) and observed a drop of -10.3\% in the average Levenshtein similarity.
These results demonstrate the performance gain of our LLM-based text reconstruction model.

\myparhead{[RQ3-2] Ablation Study of Input Reconstruction}
To validate the contribution of the reconstructed output in reconstructing the order of $K_P$, we removed the reconstructed output from the prompt of $LLM_B$, and observed that the Levenshtein similarity dropped -17.1\%.
This result confirms the validity of utilizing the reconstruct output to assist in the input reconstruction.
Finally, we completely removed the $LLM_B$, and assembled the extracted $K_P$ as the reconstructing result.
We clearly observed that Levenshtein similarity dropped -46.4\%, implying the significant
performance gain brought by $LLM_B$.
Additional results pertaining to the pure SCA are provided in the \myrefapp{additional_results}.

To demonstrate the performance gain of the side-channel attack (SCA) data, we excluded the SCA data and used only the generated output to reconstruct the input.
As illustrated in \myreftab{prefill_ablation}, we observed that the Levenshtein similarity of the pure output-based attack dropped -21\%, while the cosine similarity dropped -5.4\%.
These results show that the SCA data indeed enhances the performance, particularly at the character and token levels.
This improvement is likely due to the SCA’s ability to provide concrete tokens that narrow the search space for the target text.

\begin{table}[htb]
	\centering
	\begin{threeparttable}
		\caption{Ablation study on the SCA data.}
	\label{tab:prefill_ablation}
	\setlength{\tabcolsep}{5pt}
		\begin{tabular}{c|ccccc}
			\toprule
			\multirow{2}*{Configuration} & \multicolumn{5}{c}{Input Recovery Performance (\%)}  \\
			\cline{2-6} 
			 &  R1  & RL & LS  & $\phi$  & ASR  \\

\midrule
$LLM_B$ (SCA + Output) & 82.8& 78.5& 76.2& 96.9& 99.7\\
\midrule
$LLM_B$ (Output) & 54.8& 51.9& 55.2& 91.5& 95.8\\
			\bottomrule
		\end{tabular}
\end{threeparttable}
\end{table}

\subsection{RQ4: Framework Evaluation}
\label{sec:framework_eval}
In this research question, we validate the attack in practical scenarios, examining its viability across various LLM frameworks and computing hardware.
All victims' settings were maintained at their default values to align with the most probable real-world scenario.

\myparhead{Microbenchmark} To quickly examine whether a target framework or application is vulnerable to the side-channel, we uniformly sampled 20 samples in all the test sets to construct a microbenchmark for performance evaluation.

\myparhead{Results} 
\myreftab{frame_evalutaion} summarizes the results across different LLM inference frameworks.
We successfully attacked all 10 frameworks on CPU, 9 of them on GPU.
We failed on one GPU target called Transformers because it conducted the embedding lookup on the GPU, which cannot be probed by CPU cache side channels.

\begin{table}[!htb]
	\centering
	\begin{threeparttable}
	\caption{Evaluation on various LLM inference frameworks and different computing device. Statistics for Github starts are current until January 21, 2025.} 
	\label{tab:frame_evalutaion}
		\begin{tabular}{c|c|cc|cc}
			\toprule
		\multirow{2}*{Framework} & \multirow{2}*{\begin{tabular}{@{}c@{}} Github \\ Stars \end{tabular}} & \multicolumn{2}{c|}{CPU} & \multicolumn{2}{c}{GPU} \\
			\cline{3-6}
			&  & ${\phi}_O$ & ${\phi}_I$  &  ${\phi}_O$ & ${\phi}_I$ \\
			
			\midrule
			
			LM Studio~\cite{lmstudio} & N/A\tnote{\textdagger} & 96.6& 97.0   & 97.4& 97.3 \\
			
			\midrule
			\begin{tabular}{@{}c@{}} HuggingFace \\ Transformers ~\cite{transformers} \end{tabular} & 138k & 98.0& 74.5 &  N/A & N/A \\
			
			\cline{1-6}
			 Ollama ~\cite{ollama} & 108k & 92.0& 95.7     & 99.7& 96.1  \\
			
			\cline{1-6}
			 llama.cpp ~\cite{llama.cpp} & 71k & 99.5& 95.2  & 99.2& 97.8 \\
			
			\cline{1-6}
			 GPT4All~\cite{GPT4All} & 71k & 97.6& 95.6    & 98.9& 94.3   \\
			
			\cline{1-6}
			 LocalAI~\cite{localai} & 28k & 99.1& 97.6   & 99.0& 96.4  \\
			 
			\cline{1-6}
			 \begin{tabular}{@{}c@{}} Microsoft \\ BitNet~\cite{bitnet} \end{tabular} & 12k & 96.1& 76.0     & 98.3& 74.5  \\
			
			\cline{1-6}
			 PowerInfer~\cite{powerinfer} & 8k & 98.0& 96.5         & 98.5& 96.2 \\
			
			\cline{1-6}
			\begin{tabular}{@{}c@{}} Intel \\ IPEX-LLM~\cite{intel-ipex} \end{tabular} & 7k &  88.6& 93.8   & 96.6& 96.1  \\
			
			\cline{1-6}
			\begin{tabular}{@{}c@{}} koboldcpp~\cite{koboldcpp} \end{tabular} & 6k & 97.6& 94.9  & 99.1& 95.5 \\
						
			\bottomrule
		\end{tabular}
		\begin{tablenotes}
			\item[\textdagger] LM Studio is close-source.
		\end{tablenotes}
	\end{threeparttable}
\end{table}

\subsection{RQ5: Hardware Evaluation}
\label{sec:hardware_eval}
We evaluated the same attack across different hardware machines, utilizing the microbenchmark dataset (\myrefsec{framework_eval}) and targeting llama.cpp.
\myreftab{hardware_eval} shows that cosine similarity is at least 96.5\%, and the ASR remains 100\%.
These results are consistent across different hardware configurations, demonstrating the broad applicability of our attack methodology.

\begin{table}[!thb]
	\centering
		\caption{Evaluation on different hardware machines.}
	\label{tab:hardware_eval}
	\setlength{\tabcolsep}{5pt}
		\begin{tabular}{c|cccccccccc}
			\toprule
			\multirow{2}*{CPU} & \multicolumn{2}{c}{Output Recovery} & & \multicolumn{2}{c}{Input Recovery}  \\
			\cline{2-3} \cline{5-6}
			  & $\phi_I$ (\%) & ASR(\%) &  &  $\phi_O$(\%) & ASR(\%) \\
			
			\midrule
		Intel 14900K & 99.2& {100.0}       & & {96.5}& {100.0}  \\
			\midrule
			Intel 13900K &  99.2& 100.0    & & 97.8& 100.0        \\
			\midrule
			Intel 12700KF  & 99.3& {100.0}      & & {96.7}& {100.0}  \\
			
			\bottomrule
		\end{tabular}
\end{table}

\subsection{RQ6: Attack Examples}
In this research question, we present concrete examples to better understand the privacy leakage.
\myreffig{examples} illustrates both success and failure samples of prompts.
It is intriguing to see that the model perfectly recovered unique n-grams (such as ``freddy krueger'' and ``e5'') that were not present in the synthetic training set.
These examples highlight the potential to steal Personally Identifiable Information (PII).
Additionally, we observed that the model can maintain semantic similarity while replacing synonyms or varying grammar structures; thus, R1 and LS can drop while $\phi$ remains high in the succeeded samples.
For more detailed examples of the results from $LLM_A$ and $LLM_B$, please refer to \myrefapp{atk_examples}.

\begin{figure}[!thb]
\begin{tcolorbox}[title = Attacks on Prompts,left=1mm, right=1mm, top=1mm, bottom=0.5mm]

\setstretch{0.9}

\small
\underline{$\phi: 100\%$ \hspace{0.9em} R1: $100\%$
\hspace{0.9em} LS: $100\%$ \hspace{4.5em} \textcolor{DeepGreen}{(Succeeded)}}
\vspace{.2em}\\
\small
who played  \uwave{freddy krueger} in the 2010 nightmare on elm street?

who played \uwave{freddy krueger} in the 2010 nightmare on elm street?
\normalsize

\small
\underline{$\phi: 98\%$ \hspace{1.3em} R1: $96\%$
\hspace{1.3em} LS: $87\%$ \hspace{5.2em} \textcolor{DeepGreen}{(Succeeded)}}
\vspace{.2em}\\
\small
How can I manage my weight and \textbf{avoid gaining} excess body fat?
\vspace{.1em}

How can I manage my weight and \textcolor{BrickRed}{\textbf{avoid}} excess body fat?
\normalsize

  \myeat{
    \underline{$\phi: 100\%$ \hspace{0.5em} R1: $85\%$
    \hspace{1em} LS: $69\%$ \hspace{0.8em} Retention Key Points }
    \vspace{.2em}\\
    \small
    How long does it take to thoroughly explore the \textbf{Vatican Museum} in Rome?
    \vspace{.1em}

    How long does it take to explore the \textbf{Vatican Museum} in Rome thoroughly?
    \normalsize
    \underline{$\phi: 87\%$ \hspace{0.9em} ROUGE-1: $87\%$
    \hspace{1.1em} LS: $93\%$ \hspace{0.8em} Partial Loss}
    \vspace{.2em}\\
    \small
    who dies in the \textbf{warlots} go to birmingham?
    \vspace{.1em}
  }

\small
\underline{$\phi: 87\%$ \hspace{1.3em} R1: $78\%$
\hspace{1.3em} LS: $28\%$ \hspace{5.2em} \textcolor{DeepGreen}{(Succeeded)}}
\vspace{.2em}\\
\small
\textbf{what rank} is an  \uwave{e5} in the air force?
\vspace{.1em}

an \uwave{e5} in the air force is \textcolor{BrickRed}{\textbf{what rank}}?
\normalsize
\vspace{.1em}

\vspace{.5em}

\small
\underline{$\phi: 76\%$ \hspace{1.3em} R1: $57\%$
\hspace{1.3em} LS: $31\%$ \hspace{5.2em} \textcolor{DeepRed}{(Failed)} \hspace{1.5em}}
\vspace{.2em}\\
\small
\textbf{context:}Remy enhanced their understanding of the scientific subjects. question:What will happen to Remy?
\vspace{.1em}

What will happen to Remy \textcolor{BrickRed}{\textbf{when they have}} enhanced their understanding of scientific subjects?
\normalsize

\small
\underline{$\phi: 75\%$ \hspace{1.3em} R1: $93\%$
\hspace{1.3em} LS: $70\%$ \hspace{5.2em} \textcolor{DeepRed}{(Failed)} \hspace{1.5em}}
\vspace{.2em}\\
\small
What are the \textbf{best times} of year to visit the Grand Canyon for outdoor activities?
\vspace{.1em}

Zoeken  What \textcolor{BrickRed}{\textbf{two times}} of year are best to visit the Grand Canyon for outdoor activities?
\normalsize

\small
\underline{$\phi: 64\%$ \hspace{1.3em} R1: $47\%$
\hspace{1.3em} LS: $56\%$ \hspace{5.2em} \textcolor{DeepRed}{(Failed)}  \hspace{1.5em}}
\vspace{.2em}\\
\small
who is the national ffa president and where is he from?
\vspace{.1em}

\textcolor{BrickRed}{\textbf{headerwhere is fa president and who is he?}}
\normalsize

\end{tcolorbox}
\caption{Examples of reconstructed prompts.}
\label{fig:examples}
\end{figure}

\section{Discussion}

\subsection{Countermeasures}
We provide recommendations on mitigating the proposed attacks during the deployment and implementation of local LLM inference.

\myparhead{Disable Zero-Copy Loading} One feasible defense is to remove the zero-copy model loading that relies on \mytechno{mmap} in the LLM inference framework.
This method can eliminate the shared-memory-based cache attacks such as flush+reload and flush+flush that use \mytechno{mmap}.
However, removing the zero-copy will significantly harm performance in both temporal and spatial.
According to our evaluation results, turning off the zero-copy results in 17\% performance drops in loading latency in llama.cpp.
Moreover, the non-zero-copy version creates nearly $32\%$ extra memory overhead compared to the zero-copy version when launching two or more LLM instances for multiple users simultaneously.
Additionally, this defense cannot eliminate shared memory created by page duplication in OS~\cite{arcangeli2009increasing,singleton}.

\myparhead{Deploy Role-Based Access Control} A better mitigation is role-based access control (RBAC), which limits memory page sharing within a safe scope.
The RBAC guarantees that only designated programs are permitted to share the memory frames of the model file.
We can authenticate programs per session (through kernel hooks via eBPF) to control memory sharing access.
Unfortunately, such access control is still lacking in today's local LLM inference frameworks. 

\myparhead{Use Hardware-based Mitigation}
Intel has developed the Cache Allocation Technology (CAT) for the Xeon sever-grade CPUs, which allows software to control the LLC partition.
Based on the CAT, we can isolate the LLM inference frameworks from shared cache resources, eliminating the attacks on LLC (such as flush+reload and prime+probe).
However, CAT for LLC is typically unavailable for consumer-grade productions~\cite{guide2011intel}, on which local LLMs are mainly deployed.

\subsection{Limitation and Future Work}

\myparhead{Measurement Challenges}
The input reconstruction has relatively low accuracy in character-level and token-level metrics, which is limited by the temporal resolution of the cache attack and the parallel execution characteristic of the prefill phase.
However, the attack can still restore the input with high cosine similarity (94.8\% on average) across long text (nearly 6000 characters), resulting in an ASR of 99.1\%, i.e., the attack can successfully reveal the semantics of 99.1\% sentences under the human-evaluated cosine similarity threshold.
This high semantic similarity indicates the potential to directly extract privacy information.

\myparhead{Using Other CPU Side Channels}
The attack relies on shared-memory-based cache attacks for distinguishing between different rows in the embedding table.
Future work could explore other CPU cache attacks, such as those without shared memory, but needs to address the following \textit{challenges}:
(\romannumeral1) Conflict-based attacks (e.g., Prime+Probe) typically offer set-level spatial resolution.
Given that the embedding table size is much larger than the way size, each cache set will be mapped to multiple embedding rows, resulting in a search space consisting of many grammatically correct sentences.
To mitigate this challenge, future work could leverage LLMs to predict the sentence using inter-sentence context~\cite{remote-kla-2024}.
(\romannumeral2) Without shared memory, Address Space Layout Randomization (ASLR) and discontinuous page frame allocation can obscure the address mapping between cache sets and embedding rows.
To address this challenge, future work may require extra runtime profiling to learn the address mappings.

\myparhead{Attacking LLM Inference that Uses GPU-side Embedding}
Discrete GPUs have dedicated caches that are not coherent with the CPU's caches, making traditional CPU cache attacks infeasible for probing the discrete GPU's memory.
However, future work may explore state-of-the-art GPU cache attacks, such as Invalidate+Compare~\cite{invalidate+compare}, to monitor the memory access patterns of the token embedding.
Note that Invalidate+Compare is based on set conflicts and thus encounters the previously described \textit{challenge (\romannumeral1)}.
Despite this, Invalidate+Compare provides the additional per-set contention intensity, which could potentially reduce the search space.
Moreover, NVIDIA drivers allocate GPU page frames with less randomness, favoring physical contiguity and using consistent starting addresses~\cite{invalidate+compare}.
This deterministic behavior is beneficial for learning the address mapping between cache sets and embedding table rows.

\section{Related Work}

\myparhead{LLM Prompts or Responses Leakage}
Existing studies on breaching the confidence of LLM prompts and responses fundamentally fall under \emph{software-level} vulnerabilities, which can be divided into two categories.
One category requires access to the target model that is shared with the victim~\cite{privacy-sc-ml,chu2024-reconst-con,yang2024-prsa,hui2024-pleak,song2024earlybirdcatchesleak, zheng2024inputsnatchstealinginputllm}.
In this attack vector, adversaries must trigger requests to the model service and observe its behavior (such as response text~\cite{hui2024-pleak,privacy-sc-ml} or timing~\cite{song2024earlybirdcatchesleak,zheng2024inputsnatchstealinginputllm}) to infer the confidential information.
However, one main limitation of these types of attacks is the need for access to the victim model, which can render the malicious requests visible to the victim services.
Instead, this paper is the first to reveal both the prompts and responses via the hardware cache side-channel without directly interacting with the victim models.

Another category avoids accessing the victim model, but generally depends on intercepting network traffic or exploiting software vulnerabilities.
The keylogging attack on remote AI assistants has demonstrated the passive acquisition of model responses via intercepting encrypted network traffic~\cite{remote-kla-2024}.
This work exploits the packet-length side-channel to infer the response of online AI assistants.
Nevertheless, it relies on tapping the network communication between remote LLMs and clients, which has a threat model different from that of local LLMs.

\myparhead{Hardware Side-channel Attacks on Deep Learning} Several studies have demonstrated side-channel information leakage in deep-learning hardware, which aims to extract model structures or parameters~\cite{csinn,DeepSniffer,rev-cnn-sc,DeepTheft,cachetelephy-2020}, to infer classification labels~\cite{whispering-mlaas-2023,stealthy-inference-atk,atk-ecg-class,psca-bnn-2021}, or to infer hardware design of DNN accelerator ~\cite{dnn-ip-rev}.
However, they fundamentally target \emph{discriminative} Deep Neural Networks (DNNs). 
Instead, LLMs are generally \emph{generative} and follow the unique autoregressive paradigm.
We also found that the cache access patterns of token embedding operation leak token values, and the timing of autoregressive generation leaks token positions.
These leakages have still not been explored by previous hardware side-channel works.
To the best of our knowledge, this work is the first hardware cache side-channel eavesdropping attack on full text of LLM input and output.

\section{Conclusion}
This paper presents a novel side-channel attack to steal the model input and output text of local LLMs by leveraging the fundamental characteristics of LLM inference, including the timing of autoregressive generation and the cache access patterns of token embedding computation.
To demonstrate the feasibility, we design an eavesdropping attack framework that utilizes a new cross-modality de-noising algorithm to reconstruct the model output from the noisy cache trace.
Moreover, we fine-tune the pre-trained LLM to capture the context dependence between model input and output and reconstruct the model input from the shuffled tokens.
Finally, to overcome the lack of training data, we propose a new dataset synthesis process to obtain the training set without profiling the victim.
Our empirical evaluations across a range of mainstream LLMs demonstrated that the attack can restore high-fidelity text of model output (with an average Levenshtein similarity of 94.8\%), and reconstruct the model input with highly similar semantics (with an average cosine similarity of 98.0\%).
Our results reveal critical vulnerabilities in widely used local LLM inference frameworks (e.g., llama.cpp), highlighting the pressing necessity of improving security measures to defend against such risks.

\section{Acknowledgement}
We would like to thank the organizers, anonymous reviewers, and shepherd for their constructive comments and helpful feedback.

\section{Ethics Considerations}
All datasets utilized in our experiments are publicly available and do not contain any harmful content. Our experiments were conducted exclusively within rigorously controlled environments, ensuring no disruption to other public systems.

In adherence to best practices for vulnerability disclosure, we have reported the identified issues to all relevant software developers and are actively collaborating with them to mitigate the vulnerabilities.

\section{Open Science}
We have made our research artifact publicly available on Zenodo for permanent retrieval: \url{https://doi.org/10.5281/zenodo.15610475}. The artifact contains:

\begin{itemize}
	\item \myparhead{Complete Datasets} Including the corpus dataset, the synthesized datasets, and the collected cache traces.
	\item \myparhead{Full Source Code} Including the implementation of the proposed attack and the evaluation scripts or tools for experiment replication.
	\item \myparhead{Models} For the Llama-3.1-8B-Instruct model, full adapter weights and configuration files are provided. For the fine-tuned OpenAI's GPT-4o-mini, we are bound by proprietary restrictions that prevent sharing the model weights. As an alternative, we release the original JSONL training data files, scripts, and documents necessary to reproduce this model.
	\item \myparhead{Evaluation Data} We provide the original experiment results and anonymous survey results.
	\item \myparhead{Documentations} A comprehensive README file is available, which guides other researchers to reproduce and build upon our work.
\end{itemize}

\bibliographystyle{plain} 
\bibliography{ikwys_ref}

\appendix

\section{Dataset Construction Process}
\label{app:dataset-construction}

We used all the 2,727 testing prompts of UltraChat, and 4,425 samples in the NQ-Open, 254 prompts (12,394 tokens) in the ChatGPT-Roles, 1682 prompts in the SIQA, and 277 prompts (19,029 tokens) in the SQuAD2.
The number of total tokens is 212535.

To obtain ground-truth LLM output corresponding to different input prompts,
We send each prompt to the the general-purpose LLM (named Llama3.1-8B-Instruct for our purpose) as its input in the separated context (independent with its preceding prompts).

\section{Dataset Cleaning}
\label{app:dataset-clean}
To prevent data contamination of the training dataset from the testing samples, we dropped all the contaminated samples from the training set.
Similar to GPT-3 and PaLM~\cite{palm,gpt3}, we consider a training sample as contaminated if either its user prompts has more than 70\% 8-grams overlapping with the user prompt of testing set, or it shares at least the first 8 words with one of the testing samples.

\section{Hyperparameters of Fine-tuning}
\label{app:finetune-hyperparams}
We fine-tune OpenAi GPT-4o-mini-2024-07-18 in 3 epochs, each with a batch size of 5.
We set the learning rate multiplier as 1.8.

We fine-tune Llama-3.1-8B-Instruct in 3 epochs for $LLM_A$ and 2 epochs for $LLM_B$, each with a batch size of 2.
We set the learning rate as 0.0002 for $LLM_B$ and 0.00008 for $LLM_A$.

\section{Additional Results}
\label{app:additional_results}
\myparhead{Impact of Embedded Vector Quantization}
We evaluated the attack on other embedding vector quantization types beyond F16.
The experiment setup was aligned with \myrefsec{framework_eval}.
The attack targets the llama.cpp that uses GPU acceleration.
For Q8\_0, we skip L1 cache hits to omit extra prefetching.

\myreftab{embd_quant} shows that the quantization has a minor impact on the cosine similarity.

\begin{table}[!h]
	\centering
	\begin{threeparttable}
		\caption{Impact of embedded vector quantization.}
	\label{tab:embd_quant}
	\setlength{\tabcolsep}{1pt}
		\begin{tabular}{c|ccccc}
			\toprule
			\multirow{2}*{Quantization} & \multicolumn{2}{c}{Output Reconstruction} & & \multicolumn{2}{c}{Input Reconstruction}  \\
			\cline{2-3} \cline{5-6} 
			&  $\phi$ (\%) & ASR (\%) &  &   $\phi$ (\%) & ASR (\%) \\

			\midrule
F16 & 99.2& 100.0        & & 97.8& 100.0        \\
\midrule
Q8\_0 & 98.6& 100.0       & & 96.6& 100.0        \\
\midrule
BF16 & 99.3& 100.0       & & 96.5& 100.0        \\
\midrule
F32 & 98.7& 100.0        & & 96.2& 100.0        \\

			\bottomrule
		\end{tabular}
\end{threeparttable}
\end{table}

\myparhead{Other Operating Systems}
We evaluated the attack on other operating systems, leveraging the same model and datasets in \myrefsec{framework_eval} and fixing the victim as llama.cpp that uses GPU acceleration.
\myreftab{os_eval} shows that the attack is applicable to a range of operating systems.
\begin{table}[!h]
	\centering
	\begin{threeparttable}
		\caption{Evaluating the attack against other systems.}
	\label{tab:os_eval}
	\setlength{\tabcolsep}{1pt}
		\begin{tabular}{c|cccccccccc}
			\toprule
			\multirow{2}*{Operation System} & \multicolumn{2}{c}{Output Recovery} & & \multicolumn{2}{c}{Input Recovery}  \\
			\cline{2-3} \cline{5-6}
			&  $\phi$ (\%) & ASR (\%) &  &  $\phi$ (\%) & ASR (\%) \\
			
			\midrule
			Ubuntu 22.04 & 99.2& 100.0       & & 97.8& 100.0        \\
			\midrule
			Windows 11 & {99.4}& {100.0}      & & {95.4}& {100.0} \\
			\midrule
			Debian 12 (in Docker) & {97.0}& {100.0}       & & {96.3}& {100.0} \\
			
			\bottomrule
		\end{tabular}
\end{threeparttable}
\end{table}

\myparhead{Input Reconstruction via Pure SCA}
Experiments show that the proposed attack significantly outperforms the pure SCA, as shown in \myreftab{pure_sca}.
The setups were aligned with \myrefsec{ablation}.
\begin{table}[htb]
	\centering
	\begin{threeparttable}
		\caption{Ablation study on the SCA data.}
	\label{tab:pure_sca}
	\setlength{\tabcolsep}{3.5pt}
		\begin{tabular}{c|ccccc}
			\toprule
			\multirow{2}*{Configuration} & \multicolumn{5}{c}{Input Recovery Performance (\%)}  \\
			\cline{2-6} 
			 &  R1  & RL & LS  & $\phi$  & ASR  \\

\midrule
$LLM_B$ (SCA + Output) & 82.8& 78.5& 76.2& 96.9& 99.7\\
\midrule
Pure SCA & 50.4 &	28.7 &	29.8 &	66.1 &	29.3\\
			\bottomrule
		\end{tabular}
\end{threeparttable}
\end{table}

\section{Detailed Implementation of Flush+Reload}
\label{app:probing-pseudo-code}

In this section, we detail the implementation for obtaining cache traces using the pseudocode presented in \myreflst{pseudocode_fr}.

\definecolor{mKeyword}{RGB}{0,0,255}          
\definecolor{mString}{RGB}{160,32,240}        
\definecolor{mComment}{RGB}{34,139,34}        
\definecolor{mBackground}{RGB}{245,245,245}   
\definecolor{mNumber}{RGB}{128,128,128}       

\lstset{
  language=C++,               
  xleftmargin=2em,
  xrightmargin=1em,
  keywordstyle={\color{mKeyword}},     
  stringstyle={\color{mString}},       
  commentstyle={\color{mComment}},     
  title=\lstname,                      
  keywords={for, subroute, std, if, switch, case, return, break, int, NULL, register, struct, auto, while, asm},
  showspaces=false,                    
  showstringspaces=false,              
  showtabs=false,                      
  tabsize=2,                           
  captionpos=t,                        
  breaklines=true,                     
  numberstyle=\color{mNumber},
  numbers=left,                        
  stepnumber=1,                        
}


\begin{lstlisting}[caption={C++ pseudo code of the flush+reload attack}, label=lst:pseudocode_fr, mathescape=true, basicstyle=\footnotesize]
probe(llm_model_file) {
  //1. Achieve the shared memory (detailed error handling logics were omitted here)
  fd = open(llm_model_file, O_RDONLY);
  fstat(fd, &st_buf); // get file size
  image_ptr = mmap(NULL, st_buf.st_size, PROT_READ, MAP_PRIVATE, fd, 0);

  //2. Get the params of embedding table
  [emb_ptr, emb_stride, num_vocabs] = parse_model_file(image_ptr);
  
  //3. Generate the target address ranges
  sorted_seg = {};
  for(v = 0; v < num_vocab; v++) {
    pstart = emb_ptr + v * emb_stride[1];
    pend = pstart + emb_stride[1] - 1;
    sorted_seg = sorted_seg $\cup$ {struct AddressRange(v, pstart, pend)}
  }

  //4. Choose target addresses that satisfy the constraints to overcome prefetchers
  visited = {};
  maxpage = page_offset(emb_ptr + emb_stride[1] * num_vocab + emb_stride[1]-1) + PAGE_SIZE;
  for (cp = page_offset(emb_ptr))
    cp <= maxpage; cp += PAGE_SIZE) {
    p_center = cp + PAGE_SIZE/2;
    // find target address from ranges
    binary_search target in sorted_seg, maximizing(target.start) s.t. target.start <= p_center
    if(target is found && target->vocab not in visited) {
      visited += {target->vocab};
      target_addr[target->vocab] = p_center;
    }
  }

  //5. Avoid storing valid pointers in the array, to overcome AoP prefetchers
  for (size_t v = 0; v< num_vocab; ++v)
    ptr_offset[v] = (target_addr[v] - emb_ptr)

  //6. Collect the cache trace
  cache_trace = {};
  num_partitions = 15;
  for (t = 0; t < num_partitions; ++t) {
    // Conventional settings for the flush+reload attack
    clone(thread);
    pthread_setaffinity_np(pthread_self(), mask(t));
    sched_setscheduler(sched_get_priority_max(SCHED_OTHER));
    clflush_all_the_target_addr();

    m = ceil(num_vocabs / num_partitions)
    n = (t == num_partitions - 1) ? num_vocabs % m : m;
    while(running) {
      for (v = 0; v < n; ++v) {
        // Pseudo random permutation
        s = t * m + (v * 167 + 13) % n; // Two random coprime integers
        register p = ptr_offset[s];
        asm {
          // restore the pointer
          add emb_ptr, p
          // reload
          mfence
          rdtsc
          lfence
          mov %eax,%esi
          mov %edx,%edi
          mov (p),%ax	
          mfence
          rdtsc
          clflush (p)
        }
        timepoint = UINT64(%eax,%edx);
        cache_trace[timepoint][s] = timepoint - UINT64(%esi,%edi);
      }
    }
  }
  return cache_trace;
}
\end{lstlisting}

\section{Attack Examples and Failure Modes}
\label{app:atk_examples}

Additional concrete examples, along with an analysis of failure modes of $LLM_A$ and $LLM_B$, were made publicly available for permanent retrieval in the permanlink: \href{https://doi.org/10.5281/zenodo.15646979}{\texttt{DOI 10.5281/zenodo.15646979}}.

\end{document}